\documentclass[onecolumn,a4paper]{IEEEtran}
\usepackage{amsmath}
\usepackage{amssymb}
\usepackage{amsfonts}
\usepackage{epsfig}
\usepackage{subfigure}
\usepackage{calc}
\usepackage{color}
\usepackage[all]{xy}
\usepackage{bm}
\usepackage{algorithmicx}
\usepackage{algpseudocode}
\usepackage{algorithm}
\usepackage{enumerate}
\usepackage[noadjust]{cite}
\usepackage{caption}
\usepackage{cancel}
\usepackage{bbm}
\usepackage{url}
\usepackage{bbm}
\usepackage{breqn}

\newtheorem{defn}{Definition}
\newtheorem{thm}{Theorem}[section]
\newtheorem{cor}[thm]{Corollary}
\newtheorem{prop}{Proposition}
\newtheorem{pty}{Property}
\newtheorem{proof}{Proof}

\newtheorem{lem}[thm]{Lemma}
\newtheorem{conj}[thm]{Conjecture}
\newtheorem{constr}[thm]{Construction}
\newtheorem{note}{Remark}
\newtheorem{example}{Example}
\newcommand{\bit}{\begin{itemize}}
\newcommand{\eit}{\end{itemize}}
\newcommand{\bcor}{\begin{cor}}
\newcommand{\ecor}{\end{cor}}
\newcommand{\beq}{\begin{equation}}
\newcommand{\eeq}{\end{equation}}
\newcommand{\beqn}{\begin{equation*}}
\newcommand{\eeqn}{\end{equation*}}
\newcommand{\bea}{\begin{eqnarray}}
\newcommand{\eea}{\end{eqnarray}}
\newcommand{\bean}{\begin{eqnarray*}}
\newcommand{\eean}{\end{eqnarray*}}
\newcommand{\ben}{\begin{enumerate}}
\newcommand{\een}{\end{enumerate}}
\newcommand{\bdefn}{\begin{defn}}
\newcommand{\edefn}{\end{defn}}
\newcommand{\bnote}{\begin{note}}
\newcommand{\enote}{\end{note}}
\newcommand{\bprop}{\begin{prop}}
\newcommand{\eprop}{\end{prop}}
\newcommand{\bpty}{\begin{pty}}
\newcommand{\epty}{\end{pty}}
\newcommand{\blem}{\begin{lem}}
\newcommand{\elem}{\end{lem}}
\newcommand{\bthm}{\begin{thm}}
\newcommand{\ethm}{\end{thm}}
\newcommand{\bconj}{\begin{conj}}
\newcommand{\econj}{\end{conj}}
\newcommand{\bconstr}{\begin{constr}}
\newcommand{\econstr}{\end{constr}}
\newcommand{\bpf}{\begin{proof}}
\newcommand{\epf}{\end{proof}}
\newcommand{\bprf}{{\em Proof: }}

\newcommand{\eproof}{\hfill $\Box$}

\def\boxit#1#2{%
	\smash{\color{blue}\fboxrule=1pt\relax\fboxsep=2pt\relax%
		\llap{\rlap{\fbox{\phantom{\rule{#1}{#2}}}}~}}\ignorespaces
}

\begin{document}
\sloppy
\title{Coded Gradient Aggregation: A Tradeoff Between Communication Costs at Edge Nodes and at Helper Nodes}
\author{
\begin{minipage}{3.5in}\begin{center}
\IEEEauthorblockN{Birenjith Sasidharan} \\
\IEEEauthorblockA{Dept. of Electronics and Communication Engineering \\
Govt. Engineering College, Barton Hill, India.\\
Email: {birenjith@gecbh.ac.in} } \end{center} \end{minipage} 
\and
\begin{minipage}{3.5in} \begin{center} \IEEEauthorblockN{Anoop Thomas}  \\
\IEEEauthorblockA{School of Electrical Sciences \\ Indian Institute of Technology Bhubaneswar, India \\
Email: {anoopthomas@iitbbs.ac.in}} \end{center}
\end{minipage}}
\maketitle

\begin{abstract} 
	

The increasing amount of data generated at the edge/client nodes and the privacy concerns have resulted in learning at the edge, in which the computations are performed at edge devices and are communicated to a central node for updating the model. The edge nodes have low bandwidth and may be available only intermittently. There are helper nodes present in the network that aid the edge nodes in the communication to the server. The edge nodes communicate the local gradient to helper nodes which relay these messages to the central node after possible aggregation. Recently, schemes using repetition codes and maximum-distance-separable (MDS) codes, respectively known as Aligned MDS Coding (AMC) scheme and Aligend Repetition Coding (ARC) scheme, were proposed. It was observed that in AMC scheme the communication between edge nodes and helper nodes is optimal but with an increased cost of communication between helper and master. An upper bound on the communication cost between helpers and master was obtained. In this paper, a tradeoff between communication costs at edge nodes and helper nodes is established with the help of pyramid codes, a well-known class of locally repairable codes. The communication costs at both the helper nodes and edge nodes are exactly characterized. Using the developed technique, the exact communication cost at helper nodes can be computed for the scheme using MDS codes. In the end, we provide two improved aggregation strategies for the existing AMC and ARC schemes, yielding significant reduction in communication cost at helpers, without changing any of the code parameters. 

\end{abstract}

\section{Introduction\label{sec:intro}}

A large amount of data is getting generated at the edge nodes of the network such as mobile devices and sensors. The generated data is required to train various deep learning models to improve the performance of intelligent applications.  The edge nodes are often constrained by communication resources to share the collected data with a central server. In addition to that, the data generated may be sensitive, and sharing with a central server raises privacy concerns. Recently new techniques like federated learning and collaborative learning are considered, which enable the users to train the model at the edge node itself \cite{CELDeepPMLR, LiSTS20, ReMHP2020,ZhWZZC2020,ReTMHP2019,LiKZLCCHY2020}. This alleviates the privacy concern since the raw data is decentralized and is not shared with the central server.

In this paper, synchronous gradient descent over a decentralized data set distributed over a fixed number of edge nodes is considered. Gradient descent with centralized data is well studied and many techniques to improve the completion time of the algorithm have been proposed \cite{TaLDK2017,YeA2018,LiMMYSA2018,dutta18a,LiKMASNearOpt, HAHImpRS, AkSoSMS, RePPA2019,OzGU2019,BPAOSS, SLARTGC}.  When the data set is decentralized, the edge nodes are required to compute the local gradients which are passed to the central server to perform the global update as in federated learning. The edge nodes often have high latency, low bandwidth and may only be available intermittently. This results in straggling communication links that slow down the learning process. A technique to mitigate this problem is to consider a hierarchical setup in which reliable helper nodes are located close to the edge nodes \cite{PrRPA2020}. These helper nodes can be used by the edge nodes for efficient communication with the server node. The communication between the edge nodes and the helper nodes is unreliable due to the presence of straggling links. The helper nodes transmit the received local gradients to the central node after possible aggregation over a reliable error-free link. 

In \cite{PrRPA2020}, two schemes are considered one using repetition codes and the other using maximum-distance-separable (MDS) codes. The use of repetition code enables the reduction in communication costs at the helper nodes, however at the cost of an increased communication cost at the edge nodes. The MDS-code-approach reduces the communication cost at the edges at the expense of a larger communication cost at helpers. Inspired by this, we propose a scheme based on pyramid codes \cite{HuCL2007} which gives a tradeoff between the communication costs at edge nodes and helper nodes. In practice, the helper nodes can be shared by multiple types of edge nodes. The edge computing techniques discussed in \cite{MSNEdge} considers the use of cloudlets which are similar to the helper nodes in our architecture.  Since these nodes are shared by multiple applications, there is a possibility that the helper nodes cannot offer entire resources for transmitting gradients associated to every application. In such scenarios, the proposed scheme enables us to  adjust the communication costs at helper nodes. In the next section, we formalize the problem and present the system model.

\section{Coded Gradient Aggregation: Preliminaries\label{sec:cgd}}

\subsection{System Model\label{sec:sys}}
Gradient descent is an algorithm used in many machine learning applications to find an optimal parameter $\theta^* \, \in \,\mathbb{R}^{p\times 1}$ iteratively that minimizes a function $\sum_{x \in D} \ell(\theta,x)\,+\,\lambda R(\theta),$ where $\ell(\cdot)$ is the underlying loss function, $R(\cdot)$ denotes the regularization function, $\lambda$ is the regularization parameter, and $D$ is the data set.
The gradient $\underline{g} =\sum_{x \in D} \nabla \ell(\theta^{(t)},{x})$
is calculated at each iteration $t$ to update the parameter by the rule $\theta^{(t+1)} = \theta^{(t)} - \mu(\underline{g} + \lambda \nabla R(\theta^{(t)}))$, where $\mu$ is the learning rate. For large data sets, most of the computational effort is spent on calculating the gradient over entire dataset. 

\begin{figure}[!]
	\begin{center}
		\includegraphics[scale=0.14]{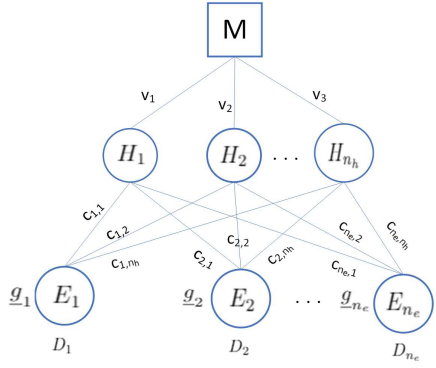}
		\caption{Hierarchical distributed learning set up with $n_e$ edge nodes and $n_h$ helper nodes.}
		\label{fig:setup}
	\end{center}
	\vspace{-0.3in}
\end{figure}

We consider the hierarchical distributed learning setup with $n_e$ edge nodes $E_1,E_2,\ldots,E_{n_e}$  and $n_h$ helper nodes $H_1, H_2, \ldots, H_{n_h}$ as illustrated in Fig. \ref{fig:setup}. For any positive integer $n$, let $[n]$ denote the set $\{1,2,\ldots,n\}$. Each edge node $E_i, i \in [n_e]$ possesses an exclusive data set $D_i$ and can compute the gradient $\underline{g}_i \in \mathbb{R}^p$ over the available data set $D_i$. The edge nodes make use of the helper nodes to communicate the partial gradients to the central master node $M$. The edge node $E_i$ splits the computed gradient $\underline{g}_i$ into $k$ vectors $\underline{g}_{i,j} \in \mathbb{R}^{(p/k)}, j \in [k]$. These gradient vectors are encoded using a linear block code (termed as {\em client code}) of length $n_h$ and dimension $k$ with generator matrix $G$ to obtain $
\underline{c}_i  =  [\underline{g}_{i,1} \ \underline{g}_{i,2} \ \cdots \underline{g}_{i,k}] \cdot G ,  \forall i \in [n_e],
$
where $\underline{c}_i = [\underline{c}_{i,1} \ \underline{c}_{i,2} \ \cdots \underline{c}_{i,n_h}]  \in \mathbb{R}^{(p/k) \times n_h}$. Each edge node $E_i, i \in [n_e]$ transmits the coded message $\underline{c}_{i,j}$ to the helper node $H_j$ for all $j \in [n_h]$. 

The communication between edge nodes and helper nodes are unreliable which is modelled by considering at most $s \in [n_h - 1]$ straggling links between each edge node and helper nodes.  Each of the helper nodes communicate to the master the observed erasure (straggling) pattern, represented by a binary vector of length $n_e$. The master constructs an $n_e \times n_h$ binary matrix referred to as the observed erasure matrix (straggling matrix), where the value $1$ in $i,j$ position indicates the straggling of the communication link between edge node $i$ and helper $j$. Let $\Omega(s)$ denote the set of all erasure matrices with $s$ straggling links per edge node. Depending on the erasure matrix $\epsilon \in \Omega(s)$, the master node instructs the helper node to encode the messages received from edge nodes to form a coded message $v_j^\epsilon \in \mathbb{R}^{(p/k) \times m_{j,\epsilon}}$. These coded messages are communicated to the master over reliable error-free links. The master makes use of the encoded messages from all the helper nodes to compute 
$
\underline{g}_D =  \left[\sum_{i=1}^{n_e} \underline{g}_{i,1} \ \sum_{i=1}^{n_e} \underline{g}_{i,2} \ \cdots \ \sum_{i=1}^{n_e} \underline{g}_{i,k}\right] .
$
The communication cost at edge nodes is quantified by the normalized communication load $C_{EH}$ where the normalization is over the local gradient vector size. Hence, we have $C_{EH}   =  \frac{(p/k) \times n_h}{p} \ = \ \frac{n_h}{k} $. The communication cost at the helper nodes depends on the straggling pattern. Hence it is quantified by considering the average helper to master communication load $C_{HM}$. The average helper to master communication load $C_{HM}$ is the total size of messages from all the helpers to master averaged over all erasure patterns $\Omega(s)$, normalized by the gradient vector size $p$. We have,
$C_{HM}  =  \ \frac{1}{|\Omega(s)|} \sum_{\epsilon \in \Omega(s)} \sum_{j=1}^{n_h} \frac{m_{j,\epsilon}}{k}. $  

For $s$ straggling links between each edge node and helper nodes, a tuple $(C_{EH}, C_{HM})$ is achievable if there exists a communication scheme between client nodes, helper nodes and the master node satisfying $C_{EH}$ and $C_{HM}$ as defined, that enables the master node to compute the gradient $\underline{g}_D$.  The set of all achievable tuples $(C_{EH}, C_{HM})$ is denoted by $\mathcal{A}$. The objective is to characterize the minimum values of the communication loads defined as follows: $$
C^*_{EH}  =  \min_{(C_{EH},C_{HM}) \in {\cal A}} C_{EH}, 
C^*_{HM}  =  \min_{(C_{EH},C_{HM}) \in {\cal A}} C_{HM}.
$$
In \cite{PrRPA2020}, it was shown that $
C^*_{EH}  = \frac{n_h}{n_h-s}$, and $
1  \leq  C^*_{HM} \ \leq \ s+1.$
The results on the optimum values of $C^*_{EH}$ and $C^*_{HM}$ are obtained from two different schemes. The bounds on the $C^*_{HM}$ are obtained by using Aligned Repetition Coding (ARC) scheme. The communication cost at edge nodes however is high for the ARC scheme. The optimum value $C^*_{EH}$ is obtained by using Aligned MDS Coding (AMC) scheme. The optimality is obtained however at an increased communication cost at the helper nodes. The results in \cite{PrRPA2020} indicates the existence of a tradeoff between the communication costs which is explored in this paper.

\subsection{Aligned MDS Coding Scheme \label{sec:amc}}
An $[n,k,d]$ code denotes a linear block code of length $n$, dimension $k$ and minimum distance $d$. In the AMC scheme, the client code is an $[n
_h, n_h - s, s+1]$ MDS code. Since the dimension of the code is $n_h -s$, $C_{EH} = \frac{n_h}{n_h - s}$. For any straggling pattern, it is guaranteed that $n_h - s$ coded messages are delivered from every edge node. Consider the naive scheme in which each helper node forwards the received messages to the master. From the MDS property, the master node is able to recover the partial gradients and compute the total gradient. This naive scheme has $C_{HM} = n_e$. In the AMC scheme, the master identifies the largest set of rows in the straggling table that match exactly. The helpers corresponding to the columns of matching rows aggregates the gradient vectors and transmit the aggregated vector to the master. For the remaining rows, the helper nodes transmit the received messages to the master. From the aggregated messages the master is able to recover an aggregated partial gradient. Using the remaining received messages the master node is able to calculate the full gradient $\underline{g}_D$. The following bound on $C_{HM}$ is obtained in \cite{PrRPA2020} by modelling the largest set of rows as the maximum occupancy of  a balls and bins problem: $C_{HM} \leq n_e - \frac{n_e}{n_h - s} + 1$.

\subsection{Aligned Repetition Coding Scheme}
Each edge node $E_i, i \in [n_e]$ partitions the computed gradient  $\underline{g}_i$ into $k$ components $\underline{g}_{i,1}, \underline{g}_{i,2}, \ldots, \underline{g}_{i,k}$, where $\underline{g}_{i,r} \in \mathbb{R}^{\frac{p}{k}}$ for $r \in [k]$. The number of components $k$ satisfies the condition that $k = \frac{n_h}{s+1}$. Each component of the gradient vector is transmitted to $s+1$ helper nodes. Since, the number of erasures is at most $s$, each partial gradient will be received by at least one helper node thus ensuring resiliency. The communication cost at the edge nodes, $C_{EH}$ for the ARC scheme is $C_{EH} = \frac{n_h}{k} = s+1$. The client code used in the ARC scheme has a generator matrix $G_{ARC} = \left[ I_{k} \ldots (s+1) \text{ times} \ldots I_k \right]$. In the ARC scheme, after obtaining the straggling pattern from helper nodes, the master takes each component and assigns the first helper node which received that particular component for transmission to the master node. Each helper node aggregates all the assigned partitions by the master and transmits an aggregated gradient vector to the master node. Note that the communication cost at each transmitting helper node is $\frac{p}{k}$. The maximum communication cost at the helper nodes arise when all the helper nodes have to transmit and hence $C_{HM} \leq \frac{n_h}{k} = s+1$. 

\subsection{Our Contributions}

\begin{figure}[h]
	\begin{center}
		\includegraphics[scale=0.5]{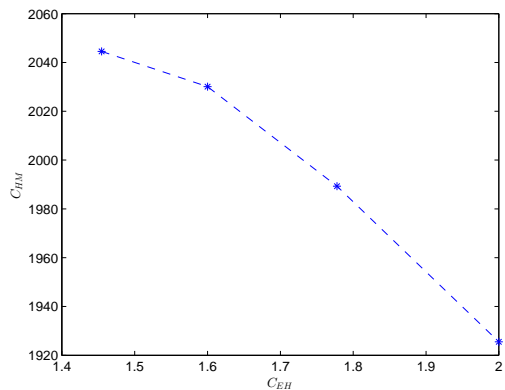}
		\caption{Tradeoff between $C_{EH}$ and $C_{HM}$ for $(n_e,n_h,s)=(2048,16,5)$}
		\label{fig:tradeoff}
	\end{center}
\end{figure}

We propose a new scheme, termed as {\em pyramid scheme}, of coded gradient aggregation based on pyramid codes, that enables to achieve a tradeoff between the communication costs at the edge nodes and at the helper nodes. In contrast to the traditional application in efficient node-repair in distributed storage, our work expands the realm of codes with locality to the new area of data communication in collaborative learning. We derive an exact expression for $C_{HM}$ and $C_{EH}$ of the proposed scheme. We also derive an exact expression for $C_{HM}$ for the existing AMC scheme. For an example parameter set $(n_e=2048,n_h=16,s=5)$, we can reduce $C_{HM}$ to $1925.5$ from $2044.5$ by increasing $C_{EH}$ from $1.45$ to $2$ as illustrated in Fig. \ref{fig:tradeoff}.
Finally, we present two improved aggregation strategies for existing AMC and ARC schemes, both achieving significant reduction in $C_{HM}$ as shown by simulations. An improvement in AMC is achieved by picking second, third, and upto $m$-th maximum when matching rows in the observed erasure matrix are identified by the master. This is in contrast to the present aggregation strategy that only considers the first maximum. An improvement in ARC is achieved by choosing helpers in a greedy manner for availing every partial gradient. 
 
\section{An Example Scheme Based on Pyramid Codes \label{sec:exmaple}}

Consider a system with $n_e = 10$ client nodes, $n_h = 8$ helper nodes with at most $s=3$ straggling links between client and helper nodes. We first construct an $[8,4,4]$ pyramid code ${\cal C}$ over $\mathbb{F}_2^3$ and use it as the client code. Starting from the systematic generator matrix $\tilde{G}_0$ of $[7,4,4]$ Reed-Solomon code over $\mathbb{F}_2^3$
\bean
\tilde{G}_{0} & = & \left[ \begin{array}{ccccccc}
	1 & 0 & 0 & 0 & a_{11} & a_{12} & a_{13} \\   
	0 & 1 & 0 & 0 & a_{21} & a_{22} & a_{23} \\   
	0 & 0 & 1 & 0 & a_{31} & a_{32} & a_{33} \\   
	0 & 0 & 0 & 1 & a_{41} & a_{42} & a_{43}   			
\end{array} \right] 
\eean
we can construct the generator matrix $G_0$ of $[8,4,4]$ pyramid code ${\cal C}_0$ as
\bean
G_0 & = & \left[ \begin{array}{cccccccc}
	1 & 0 & 0 & 0 & a_{11} & 0 & a_{12} & a_{13} \\   
	0 & 1 & 0 & 0 & a_{21} & 0 & a_{22} & a_{23} \\   
	0 & 0 & 1 & 0 & 0 & a_{31} & a_{32} & a_{33} \\   
	0 & 0 & 0 & 1 & 0 & a_{41} & a_{42} & a_{43}   			
\end{array} \right].
\eean
Every real gradient vector can be quantized so that it is represented using a finite number of bits. In that case, it is possible to encode it using a client-code over finite-field. There is a parallel line of work \cite{AlGLTV2017,AcDFS2019,ShCEPC2020} that attempts to establish that such quantizations do not affect the performance of algorithms making use of the gradients.  Every client node splits the local gradient vector into $4$ sub-vectors each of length $(p/4)$ and use the same code ${\cal C}_0$ to encode the gradient vector to produce a vector of length $(p/4)\cdot 8=2p$. Thus we obtain a $(10 \times 8)$ codeword array with every entry being a $(p/4)$-long vector. It is easy to find that $C_{EH} = \frac{8}{4} = 2$. The code ${\cal C}_0$ has two local single parity check codes of dimension $2$ with supports $L_1 = \{1,2,5\}$ and $L_2 = \{3,4,6\}$. The global parities are over the support $P=\{7, 8\}$. Suppose the observed erasure matrix is as depicted by cross-marks in Fig.~\ref{fig:erasureexample}. Thus the observed erasure matrix is given by
\bean
\mathcal{M}_E & = & \left[ \begin{array}{cccccccc}
	0 & 1 & 0 & 1 & 0 & 0 & 1 & 0 \\   
	0 & 1 & 0 & 1 & 0 & 0 & 0 & 1 \\   
	0 & 1 & 0 & 1 & 0 & 0 & 0 & 1 \\   
	1 & 0 & 1 & 1 & 0 & 0 & 0 & 0 \\   
	0 & 1 & 1 & 1 & 0 & 0 & 0 & 0 \\   
	0 & 0 & 1 & 1 & 1 & 0 & 0 & 0 \\   
	0 & 0 & 1 & 1 & 1 & 0 & 0 & 0 \\   
	0 & 0 & 0 & 1 & 0 & 0 & 1 & 1 \\   
	0 & 0 & 1 & 0 & 0 & 0 & 1 & 1 \\   
	0 & 0 & 1 & 1 & 0 & 0 & 0 & 1 
\end{array} \right].
\eean
\begin{figure}
	\centering 
	\scalebox{.9}{
		\begin{tabular}{|c|c|c|c|c|c|c|c|c|} 
			\hline
			& $H_1$ & $H_2$ & $H_3$ & $H_4$ & $H_5$ & $H_6$ & $H_7$ & $H_8$ \\
			\hline
			$E_1$ & $c_{11}$ &  \xcancel{$c_{12}$} &  $c_{13}$ &  \xcancel{$c_{14}$} &  $c_{15}$ & $c_{16}$ & \xcancel{$c_{17}$} & $c_{18}$  \\
			\hline 
			$E_2$ & $c_{21}$ &  \xcancel{$c_{22}$} &  $c_{23}$ &  \xcancel{$c_{24}$} &  $c_{25}$ & $c_{26}$ & $c_{27}$ & \xcancel{$c_{28}$}   \\
			\hline
			$E_3$ & \boxit{.2in}{.38in} $c_{31}$ &  \xcancel{$c_{32}$} &  \boxit{.2in}{.38in}$c_{33}$ &  \xcancel{$c_{34}$} &  \boxit{.2in}{.38in}$c_{35}$ & \boxit{.2in}{.38in}$c_{36}$ & $c_{37}$ & \xcancel{$c_{38}$} \\        
			\hline 
			$E_4$ & \xcancel{$c_{41}$} &  \boxit{.2in}{.07in}$c_{42}$ &  \xcancel{$c_{43}$} &  \xcancel{$c_{44}$} & \boxit{.2in}{.07in}$c_{45}$ & $c_{46}$ & $c_{47}$ & $c_{48}$  \\
			\hline 
			$E_5$ &  \boxit{.2in}{.07in}$c_{51}$ &  \xcancel{$c_{52}$} &  \xcancel{$c_{53}$} &  \xcancel{$c_{54}$} & \boxit{.2in}{.07in} $c_{55}$ & $c_{56}$ & $c_{57}$ & $c_{58}$   \\
			\hline 
			$E_6$ & \boxit{.2in}{.07in}$c_{61}$ & \boxit{.2in}{.07in} $c_{62}$ &  \xcancel{$c_{63}$} &  \xcancel{$c_{64}$} &  \xcancel{$c_{65}$} & $c_{66}$ & $c_{67}$ & $c_{68}$   \\       \hline 
			$E_7$ & \boxit{.2in}{.07in}$c_{71}$ & \boxit{.2in}{.07in} $c_{72}$ &  \xcancel{$c_{73}$} &  \xcancel{$c_{74}$} &  \xcancel{$c_{75}$} & $c_{76}$ & \boxit{.2in}{.59in} $c_{77}$ & \boxit{.2in}{.59in} $c_{78}$   \\       \hline 
			$E_8$ & \boxit{.2in}{.07in}$c_{81}$ &  \boxit{.2in}{.07in}$c_{82}$ &  \boxit{.2in}{.07in}$c_{83}$ &  \xcancel{$c_{84}$} &  $c_{85}$ & \boxit{.2in}{.07in}$c_{86}$ & \xcancel{$c_{87}$} & \xcancel{$c_{88}$}   \\       \hline 
			$E_9$ &  \boxit{.2in}{.07in}$c_{91}$ &  \boxit{.2in}{.07in}$c_{92}$ &  \xcancel{$c_{93}$} &  \boxit{.2in}{.07in}$c_{94}$ &  $c_{95}$ & \boxit{.2in}{.07in}$c_{96}$ & \xcancel{$c_{97}$} & \xcancel{$c_{98}$}   \\       \hline 
			$E_{10}$ &  \boxit{.25in}{.07in}$c_{10,1}$ &  \boxit{.25in}{.07in}$c_{10,2}$ &  \xcancel{$c_{10,3}$} &  \xcancel{$c_{10,4}$} &  $c_{10,5}$ & \boxit{.25in}{.07in}$c_{10,6}$ & \boxit{.25in}{.07in}$c_{10,7}$ & \xcancel{$c_{10,8}$}  \\     \hline 
	\end{tabular} }
	\caption{\footnotesize Aggregation strategy for $(n_e=10,n_h=8,s=3)$ using $[8,4,4]$ pyramid code with $2$ local codes.}
	\label{fig:erasureexample}
\end{figure} 
Given an erasure pattern, a local code is said to be unaffected by the erasure pattern, if every symbol in the local code can be decoded. Otherwise, the local code is said to be overwhelmed by the erasure pattern. The coded symbols are aggregated (summed up) at helper nodes before transmitting to the master. A set of columns $T$ is picked such that 
\bit
\item they do not associate to global parities if no local code is overwhelmed, or else,
\item they associate to overwhelmed local codes and global parities,
\eit
and hold unerased information symbols. Restricted to these columns, set of rows $\Psi$ with a matching erasure pattern is picked. Symbols in every column of $T$ are summed up over $\Psi$. Let us look at how it works as applied to the given example. 
\ben
\item Over the rows $E1-E3$, we aggregate symbols over the columns $\{1,2,\ldots, 6\}$. This is because neither $L_1$ nor $L_2$ are overwhelmed. Thus we choose columns in $L_1 \cup L_2$ as candidate columns for aggregation. The information set within these columns is $\{1,3,5,6\}$. Therefore the helpers $H_j,j=1,3,5,6$ send $c_{1j}+c_{2j}+c_{3j}$ to the master.
\item Over the rows $E4-E7$, the local code $L_2$ is overwhelmed. So we choose $L_2 \cup P = \{3,4,6,7,8 \}$ as columns for aggregation. Symbols belonging to unaffected local code $L_1$ are sent without aggregation. Once $L_1$ is completely decoded in every row, then symbols in columns $\{1,2\}$ are available at every row $E4-E7$. Since $\{1,2,7,8\}$ is a information set, symbols are aggregated and sent for columns $\{7,8\}$. Thus helpers $H_j$, $j=7,8$ send aggregated symbols $c_{4j}+c_{5j}+c_{6j}+c_{7j}$.
\item For $E8$, both the local codes are unaffected. Hence aggregation will be done over $L_1 \cup L_2$. However there are no other row with a matching erasure pattern. So the symbols from this row will be sent without aggregation. 
\item The case of $E9$ is exactly similar to the case of $E8$.
\item For $E10$, $L_2$ is an overwhelmed code. Thus aggregation will be attempted over columns $L_2 \cup P$. However, there are no rows left out. For this reason, symbols from $E10$ will be sent without aggregation. 
\een 
The above described aggregation is illustrated in Figure \ref{fig:erasureexample}.

\section{The Pyramid Scheme \label{sec:pyramid}}

The example described in Sec.~\ref{sec:exmaple} can be generalized to arrive at a new scheme for coded aggregation of gradient. In this section, we present the new scheme, referred to as {\em pyramid scheme}, for a fixed parameter set $(n_e,n_h,s)$. Let $t \geq 2$ be an integer parameter. A pyramid scheme is a pair $({\cal C}_t, {\cal A}_t)$ indexed by $t$, where ${\cal C}_t$ is a pyramid code and ${\cal A}_t$ is the associated aggregation strategy. For the degenerate case of $t=1$, the scheme reduces to the AMC scheme.  In the following, we describe the scheme and characterize its communication cost-pair $(C_{\text{EH},\text{Pyr}}(t), C_{\text{HM},\text{Pyr}}(t))$ for every possible value of $t$. 

\subsection{Construction of a Set of Pyramid Codes}

Let the parameters be $(n_e,n_h,s)$. We set $k_1 = n_h-s$ and define $k_t = k_1 -t +1, \ 2 \leq t \leq \lfloor \frac{k_1+1}{3} \rfloor$. Construct a $[k_t+s,k_t,s+1]$-MDS code over $\mathbb{F}_q$ with generator matrix
\bean
G_{t, \text{\scriptsize MDS}} & = & [I_{k_t} \ P] \\
& = & \left[ \begin{array}{ccccc}
	I_{k_t} & \underline{p}_{1} & \underline{p}_{2} & \cdots & \underline{p}_{s} 
\end{array} \right] 
\eean
where $I_{k_t}$ is an identity matrix of size $k_t$ and $\underline{p}_i, 1 \leq i \leq s$ are columns of the $(k_t \times s)$ Cauchy matrix. The field size $q$ may be suitably chosen. We transform $G_{t, \text{\scriptsize MDS}}$ to derive the generator matrix $G_t$ of a pyramid code ${\cal C}_t$ with blocklength $n_h$ and dimenstion $k_t$.  First, we write
\bean
k_t & = & ta + b, \ 0 \leq b < t \\
&  = & b(a+1) + (t-b)a
\eean
and split the parity vector $\underline{p}_1$ into $t$ vectors $\underline{p}_{11}, \underline{p}_{12}, \ldots, \underline{p}_{1t}$ such that the first $b$ vectors $\underline{p}_{11}, \underline{p}_{12} , \ldots  \underline{p}_{1b}$ belong to $\mathbb{F}_q^{a+1}$ and the remaining $(t-b)$ vectors $\underline{p}_{1,b+1}, \underline{p}_{1,b+2} , \ldots  , \underline{p}_{1t}$ belong to $\mathbb{F}_q^{a}$. Then we construct $G_{t} = [I_{k_t} \ P]$ where $P$ is a $k_t \times (n_h-k_t)$ matrix 
\bean
P & = & \left[ \begin{array}{ccccccccc}
	 \underline{p}_{11} & & & & & & & &  \\
	  & \underline{p}_{12} & & & & & & &  \\
	  & & \ddots & & & &  &  &  \\
	  & & & \underline{p}_{1b} & & &  \underline{p}_{2} & \cdots &  \underline{p}_{s} \\
	  & & & & \underline{p}_{1,b+1} & &  & &  \\
	  & & & & \ddots &  & & &  \\
	  & & & & & \underline{p}_{1t} & & &
\end{array} \right] 
\eean
We put in place the required definitions and subsequently characterize the parameters of ${\cal C}_t$.
\bdefn \ben 
\item An symbol $c_i, \ 1 \leq i \leq n$ an $[n,k]$ code ${\cal C}$ is said to have locality $r$ if there is a punctured code $C_i$ of dimension $r$ containing $c_i$.  
\item  An $[n,k]$ code ${\cal C}$ is said to be a code with locality $r$ if every symbol has locality $r$. If every symbol $c_i$ has locality $r_i$, and $r_i$ is not necessarily equal to $r_j$ for $i \neq j$, then the code is said to be a code with unequal locality.
\item The vector $(\kappa_1, \kappa_2, \ldots, \kappa_r)$ is said to be the locality profile of an $[n,k]$ code ${\cal C}$ with unequal locality if
\bit
\item $\kappa_1 + \kappa_2 + \cdots + \kappa_r = k$
\item there are $\kappa_j$ information symbols in ${\cal C}$ with locality $j$, $1 \leq j \leq r$.
\eit 
\een 
\edefn
The code ${\cal C}_t$ generated by $G_t$ is a pyramid code \cite{HuCL2007} with $t$ local codes. It is a code with unequl locality \cite{GoHSY2012} \cite{KaS2016} with every local code a single-parity-check code. 
\blem \label{lem:code} The code ${\cal C}_t, \ 2 \leq t \leq \lfloor \frac{k_1+1}{3} \rfloor$ is an $[n_h, k_t, s+1]$ code. It can thus correct any $s$ erasures.
\elem
\bprf The block length of ${\cal C}_t$ can easily be computed as $k_t+t+(s-1) = k_1+s = n_h$. By construction, the locality profile of the code 
\bean
\underline{\kappa} &  =  & (\kappa_1, \kappa_2,\cdots, \kappa_a, \kappa_{a+1}) \ = \ (0,0,\cdots, 0, (t-b)a,b(a+1)) .
\eean 
By Theorem~$1$ in \cite{KaS2016}, the minimum distance of the code 
\bean
d & \leq & n-k - \sum_{j=1}^{a+1} \left\lceil \frac{\kappa_j}{j} \right\rceil + 2 \\
& = & n_h-k_t - \left\lceil \frac{(t-b)a}{a} \right\rceil - \left\lceil \frac{b(a+1)}{a+1} \right\rceil + 2 \\
& = & n_h-k_1+1 \ = \ s+1. 
\eean
It is shown in \cite{KaS2016} that pyramid codes acheive the upper bound on minimum distance. This completes the proof.
\eproof 

Furthermore, we can explicitly identify the supports of all $t$ local codes as $L_i \ = \ \{(i-1)(a+1)+1, (i-1)(a+1)+2, \ldots, i(a+1),k_t+i \},  1 \leq i \leq b$ and $L_i = \{(i-1)a+b+1,(i-1)a+b+2,\ldots, ia+b,,k_t+i\},  b+1 \leq i \leq t$. Then we define $Q = [n] \setminus \cup_{i \in [t]} L_i$ as the support of all global parities that do not participate in any local code. We shall also define $k_t(i), 1 \leq i \leq t$ as the dimension of the local code with support $L_i$. Clearly $\sum_{i=1}^t k_t(i) = k_t$. In the construction of ${\cal C}_t$, we do not wish to end up with a local code of trivial dimension $1$. Thus we shall require $ta + b = k_1-t+1$ with $a\geq 2$, $0\leq b < t$ and thereby it follows that
\bean
t & = & \frac{k_1+1-b}{a+1} \ \leq \ \left\lfloor \frac{k_1+1}{3} \right\rfloor .
\eean 

\subsection{A Classification of Erasure Patterns}

As discussed in Sec.~\ref{sec:sys}, an erasure (failure) pattern of links connecting edge clients to helpers can be represented as a binary vector of length $n_h$ with Hamming weight $s$. Let ${\cal E}$ denote the set of all erasure patterns, and we have $|{\cal E}| = {n_h \choose s}$. Every edge node will be subjected to an erasure pattern, and there are a total of $n_e$ erasure patterns witnessed by the system. The aggregation strategy ${\cal A}_t$ adopted by the helpers (as prescribed by the master) depends on these erasure patterns. Here we shall discuss a partition of ${\cal E}$ based on ${\cal C}_t$ that will become necessary in the subsequent section to describe ${\cal A}_t$.  

For any binary vector $\underline{x} = (x_1,x_2,\ldots, x_n)$, we define $w_H(\underline{x})$ as the Hamming weight of $\underline{x}$, and $\textsl{Supp}(\underline{x})$ as the support of $\underline{x}$ i.e., $\{ i \mid x_i \neq 0\}$. For the code ${\cal C}_t$, we say a local code or  its support $L_i, 1 \leq i \leq t$ (by overloading of definition) is {\em overwhelmed} by an erasure pattern $\underline{e} \in {\cal E}$ if $\textsl{Supp}(\underline{e}) \cap L_i > 1$. Otherwise, if $\textsl{Supp}(\underline{e}) \cap L_i \leq 1$, we say the local code or $L_i$ is {\em unaffected} by $\underline{e}$. For a given erasure pattern, let $\tau$ denote the number of unaffected local codes, and $i_1 < i_2 < \cdots < i_{\tau}$ denote the ordered indices of unaffected local codes. Similarly, $j_1 < j_2 \cdots < j_{t-\tau}$ denote the ordered indices of overwhelmed local codes. We define $f_i = |\textsl{Supp}(\underline{e}) \cap L_{i}|$ as the number of erasures within a local code $L_i, \  1\leq i \leq t$, and $f_{t+1}=|\textsl{Supp}(\underline{e}) \cap Q|$ as the number of erasures within global parities. The vector $\underline{f}=(f_1,f_2,\ldots,f_{t+1})$ is referred to as the {\em code-erasure pattern}. Let $\lambda_i = |L_i|, \ 1\leq i \leq t$ denote the size of supports of local codes, and $\lambda_{t+1}=|Q|=(s-1)$ denote the number of global parities. Then there are 
\bean
N(\underline{f}) & = & \prod_{i=1}^{t+1} {\lambda_i \choose f_i}
\eean
erasure patterns associated with a particular code-erasure pattern.

Let $\underline{u} \in \mathbb{F}_2^t$ and $\underline{v} \in \mathbb{F}_2^{t-w_H(\underline{u})}$. We define a subset $S_{(\underline{u},\underline{v})} \subset {\cal E}$ indexed by the tuple $(\underline{u},\underline{v})$ as follows. An erasure pattern $\underline{e}$ belongs to $S_{(\underline{u},\underline{v})}$ if  the following two conditions hold:
\ben
\item[(a)] For every $i \in \textsl{Supp}(\underline{u})$, $L_i$ is overwhelmed by $\underline{e}$, thereby implying $\tau=t-w_H(\underline{u})$,
\item[(b)] If ${i_1},i_2,\ldots {i_{\tau}}$ are the ordered indices of unaffected local codes, then $(f_{i_1},f_{i_2},\ldots, f_{i_{\tau}})=\underline{v}$.
\een
This means that (a) $\underline{u}$ determines which all local codes are overwhelmed by $\underline{e}$, and that (b)  $\underline{v}$ determines the number of erasures (either $0$ or $1$) within unaffected local codes. We say that two erasure patterns in ${\cal E}$ are of {\em same type}, if they belong to the same subset $S_{(\underline{u},\underline{v})}$. And we write $\textsl{type}(\underline{e}) = (\underline{u}, \underline{v})$ if $\underline{e} \in S_{(\underline{u},\underline{v})}$. The size of $S_{(\underline{u},\underline{v})}$ can be counted as:
\bean
|S_{(\underline{u},\underline{v})}| & = & {n_h-\sum_{\ell \notin \{ i_1,i_2,\ldots i_{\tau} \}}\lambda_{\ell} \choose s-w_H(\underline{v})} \prod_{\ell =1}^{\tau} {\lambda_{i_{\ell}} \choose v_{\ell}} -  \vspace{0.5in}\sum_{\substack{\underline{x} \in \mathbb{F}_2^t \\ (x_{i_1},x_{i_2},\ldots, x_{i_{\tau}})=\underline{v}}}   {s-1 \choose s-w_H(\underline{x})} \prod_{\ell =1}^{t} {\lambda_{\ell} \choose x_{\ell}} 
\eean
Next, we shall define a relation $\simeq$ within $S_{(\underline{u},\underline{v})}$. Define $A(\underline{u}) = \cup_{i\in \textsl{Supp}(\underline{u})} L_i \cup Q$ as the support of all overwhelmed local codes combined with that of global parities. Define  $B(\underline{u}) = \cup_{i\in [t]\setminus \textsl{Supp}(\underline{u})} L_i$ as the support of unaffected local codes. Let $\underline{e}_1, \underline{e}_2 \in {\cal E}$. We say that $\underline{e}_1$ is {\em equivalent} to $\underline{e}_2$ or $\underline{e}_1 \simeq \underline{e}_2$ if they are of same type and 
\bean
\textsl{Supp}(\underline{e}_1) \cap A(\underline{u}) & = &  \textsl{Supp}(\underline{e}_2) \cap A(\underline{u}) \text{ if } \underline{u} \neq \underline{0}, \\
\textsl{Supp}(\underline{e}_1) \cap B(\underline{u}) & = &  \textsl{Supp}(\underline{e}_2) \cap B(\underline{u}) \text{ if } \underline{u} = \underline{0} .
\eean
In otherwords, equivalent erasure patterns remain the same at every location except possibly within unaffected local codes, and their code-erasure pattern remains the same within unaffected local codes. It is easy to verify that $\simeq$ is an equivalence relation, and hence partitions $S_{(\underline{u},\underline{v})}$ into equivalence classes. The set of equivalence classes is denoted by ${\cal B}=S_{(\underline{u},\underline{v})}/\simeq$, and we define $\mu_{(\underline{u},\underline{v})} = |{\cal B}|$. Whenever the type $(\underline{u},\underline{v})$ is evident from the context, we simply write $\mu$ in place of $\mu_{(\underline{u},\underline{v})}$. The equivalence classes within ${\cal B}$ are denoted by $B_1(\underline{u},\underline{v}), B_2(\underline{u},\underline{v}),\ldots, B_{\mu}{(\underline{u},\underline{v})}$. The size of every equivalence class $B_j(\underline{u},\underline{v})$, $1 \leq j \leq \mu$ is the same and  we define 
\bean
\beta(\underline{u},\underline{v}) & = & |B_j(\underline{u},\underline{v})| \\
& = & \left\{ \begin{array}{ll} \prod_{\ell \in [t] \setminus \textsl{Supp}(\underline{u})} {\lambda_{\ell} \choose f_{\ell}} & \underline{u} \neq \underline{0} \\
{\lambda_{t+1} \choose f_{t+1}} & \underline{u} = \underline{0} \end{array} \right. 
\eean
where $(f_1,f_2,\ldots, f_{t+1})$ is the code-erasure pattern of $\underline{e}$. We refer to $\beta(\underline{e})$ as the {\em bunching factor} of every $\underline{e} \in B_j(\underline{u},\underline{v})$, for reasons that will become apparent in later subsections. Since $(f_{i_1},f_{i_2},\ldots, f_{i_u})$ remains the same for every erasure pattern in $S_{(\underline{u},\underline{v})}$, the bunching factor too remains the same for them. Thus $\simeq$ partitions $S_{(\underline{u},\underline{v})}$ into equivalence classes of the same size. An example partition of ${\cal E}$ is illustrated in Fig.~\ref{fig:eg1}.

\begin{figure}
	\bean 
	\footnotesize 
	\centering
	\begin{array}{||c|c|c|c|c||} 
		\hline \hline
		\text{Type } (\underline{u},\underline{v}) & |S_{(\underline{u},\underline{v})}| & \underline{f}=(f_1,f_2,f_3)  & N(\underline{f}) & \beta(\underline{u},\underline{v}) \\
		\hline \hline 
		(00,00) & 0 & - & 0 & - \\
		\hline
		(00,01) & 6 & (0,1,4) & 6 & 1 \\
		\hline
		(00,10) & 6 & (1,0,4) & 6 & 1 \\
		\hline
		(00,11) & 144 & (1,1,3) & 144 & 4 \\
		\hline 
		(01,0) &  246 & (0,2,3) & 60 & 1 \\
		\cline{3-5} 
		& & (0,3,2) & 120 & 1 \\ 
		\cline{3-5} 
		&& (0,4,1) & 60 & 1 \\ 
		\cline{3-5} 
		&& (0,5,0) & 6 & 1 \\ 
		\hline 
		(01,1) & 1110 & (1,2,2) & 540 & 6 \\
		\cline{3-5} 
		&& (1,3,1) & 480 & 6 \\ 
		\cline{3-5} 
		&& (1,4,0) & 90 & 6 \\ 
		\hline 
		(10,0) &  246 & (2,0,3) & 60 & 1 \\
		\cline{3-5} 
		&& (3,0,2) & 120 & 1 \\ 
		\cline{3-5} 
		&& (4,0,1) & 60 & 1 \\ 
		\cline{3-5} 
		&& (5,0,0) & 6 & 1 \\ 
		\hline 
		(10,1) & 1110 & (2,1,2) & 540 & 6 \\
		\cline{3-5}
		&& (3,1,1) & 480 & 6 \\ 
		\cline{3-5}
		&& (4,1,0) & 90 & 6 \\ 
		\hline 
		(11,-) & 1500 & (2,2,1) & 900 & 1 \\
		\cline{3-5} 
		& & (2,3,0) & 300 & 1 \\
		\cline{3-5}
		& & (3,2,0) & 300 & 1 \\
		\hline \hline 
	\end{array}
	\eean
	\caption{Classification of erasure patterns for a $[16,10,6]$ pyramid code with $t=2$, $(\lambda_1,\lambda_2,\lambda_3)=(6,6,4)$. \label{fig:eg1}}
\end{figure}

\subsection{The Aggregation Strategy ${\cal A}_t$}

As discussed in Sec.~\ref{sec:sys}, the codeword transmitted from all edges is an $(n_e \times n_h)$ matrix over $\mathbb{F}_q$
\bean
C & = & \left[ \begin{array}{cccc}
	c_{11} & c_{12} & \cdots & c_{1,n_h} \\
	c_{21} & c_{22} & \cdots & c_{2,n_h} \\
	\vdots & \vdots & \ddots & \vdots \\
	c_{n_e,1} & c_{n_e,2} & \cdots &c_{n_e,n_h}
\end{array} \right]
\eean
All $n_e$ erasure patterns together can be represented as a matrix ${\cal M}_E$ over $\mathbb{F}_2$.
\bean
{\cal M}_E & = & \left[ \begin{array}{cccc}
	\epsilon_{11} & \epsilon_{12} & \cdots & \epsilon_{1,n_h} \\
	\epsilon_{21} & \epsilon_{22} & \cdots & \epsilon_{2,n_h} \\
	\vdots & \vdots & \ddots & \vdots \\
	\epsilon_{n_e,1} & \epsilon_{n_e,2} & \cdots &\epsilon_{n_e,n_h}
\end{array} \right] 
\ := \ \left[ \begin{array}{c}
	\underline{\epsilon}_{1} \\
	\underline{\epsilon}_{2} \\
	\vdots \\
	\underline{\epsilon}_{n_e} 
\end{array} \right] .
\eean
We refer to ${\cal M}_E$ as the observed erasure matrix. The code symbol $c_{ij}$ is unerased if $\epsilon_{ij}=0$, and erased if $\epsilon_{ij}=1$. Here $\underline{\epsilon}_i , \ 1 \leq i \leq n_e$ is the erasure pattern observed by the $i$-th edge. It is possible that an erasure pattern can repeat at multiple edge-clients, and thus $E = \{ \underline{\epsilon}_1, \underline{\epsilon}_2, \ldots, \underline{\epsilon}_{n_e} \}$ is a multi-set of observed erasure patterns. 

Our scheme is inspired by the AMC scheme proposed in \cite{PrRPA2020}. In AMC, code symbols from different rows of the codeword array are ``aggregated'' and transmitted to reduce the communication cost from helpers to the master. In our approach, we will carry out aggregation of symbols leveraging upon the local-decoding capabilities of ${\cal C}_t$. The aggregation strategy will take into account the distribution of overwhelmed and unaffected local codes. We begin with formalizing what is meant by aggregation of code symbols.

\bdefn Let the code ${\cal C}_t$ and the observed erasure matrix ${\cal M}_E$ be given. We say that the code-symbols are {\em aggregated} over a set of rows $R \subset [n_e]$ and a set of columns $T \subset [n_h]$ if the following operations are executed in order:
\ben
\item[(a)] Identify a minimal subset $I_{R,\overline{T}}$ of $C_{R,\overline{T}}=\{ c_{i,j} \mid j \in R, i \in [n]\setminus T\}$ such that symbols in $I_{R,\overline{T}}$ are unerased, and it is possible to recover $C_{R,\overline{T}}$ given the knowledge of $I_{R,\overline{T}}$. 
\item[(b)] Identify subsets $T_1 \subseteq T$, $\overline{T}_1 \subset [n]\setminus T$ such that (i) $ T_1 \cup \overline{T}_1$ is an information set of ${\cal C}_t$, and (ii) $c_{i,j}$ is not erased for every $j \in T_1$ and every $i \in R$.
\item[(c)] Compute 
\bean
d_{j} & = & \sum_{i\in R} c_{ij}, \ \ j \in T_1
\eean 
and generate the set $D = \{d_j \mid j \in T_1\} \cup I_{R,\overline{T}}$. 
\een
We refer to $D$ as the {\em aggregated set of code symbols} over a set of rows $R$ and a set of columns $T$. 	
\edefn
In the following, we describe the aggregation strategy ${\cal A}_t$ associated with the pyramid code ${\cal C}_t, \ 2 \leq t \leq \lfloor \frac{n_h-s+1}{3}\rfloor$. The strategy takes in the codeword $C$ and the observed erasure matrix ${\cal M}_E$ (or the multi-set $E$) as inputs.
\ben
\item Let $\{(\underline{u}_j,\underline{v}_j) \mid j=1,2,\ldots, j_{\text{max}} \}$ be the set of types of erasure patterns in $E$. Let us write $\mu_j = \mu(\underline{u}_j,\underline{v}_j)$. Let $\Psi_{j,\ell} = \{ i \mid \textsl{type}(\underline{e}_i) = (\underline{u}_j,\underline{v}_j), \ \underline{e}_i \in B_j(\underline{u}_j,\underline{v}_j) \}, \ j=1,2,\ldots, j_{\text{max}}, \ell=1,2,\ldots, \mu_j$. 
\item For every $j=1,2,\ldots, j_{\text{max}}$, do the following:
\ben \item[(a)] compute $\ell^*(j)=\arg\max_{\ell=1,2,\ldots,\mu_j} |\Psi_{j,\ell}|$
\item[(b)] Let $T(\underline{u}_j) = A(\underline{u}_j)$ if $\underline{u}_j \neq \underline{0}$, or else $T(\underline{u}_j) = B(\underline{u}_j)$. Aggregate code symbols over set of rows $\Psi_{j,\ell^*(j)}$ and columns $T(\underline{u}_j)$. Let $D_j$ be the aggregated set of symbols.  
\een
\item For every row $\ell \in [n_e] \setminus \cup_{j=1}^{j_{\text{max}}}\Psi_{j,\ell^*(j)}$, identify a set of unerased information symbols $I_{\ell}$. Let $D_{j_{\text{max}}+1} = \cup_{\ell } I_{\ell}$.
\item The master instructs helpers to transmit $D_{\text{ag}}=\cup_{j=1}^{j_{\text{max}}+1} D_{j}$.
\een

\subsection{Calculation of $C_{\text{EH}}$ and $C_{\text{HM}}$}

By Lemma~\ref{lem:code}, the code ${\cal C}_t$ has dimension $k_t=n_h-s-t+1$. It follows that the communication cost by edge-nodes for the pyramid scheme $({\cal C}_t,{\cal A}_t)$ is given by:
\bea \label{eq:cehpyr}
C_{\text{EH},\text{Pyr}}(t) &  =  & \frac{n_h}{n_h-s-t+1}
\eea

We assume that the observed erasure matrix is a uniform random variable in $\mathbb{F}_2^{n_e\times n_h}$. Since the aggregation strategy depends on the erasure multi-set $E$, the size of aggregated set of symbols $D$ is a random variable. Thus $C_{\text{HM}}$ is the expected value of the random variable $|D_{\text{ag}}|/k_t$:
\bea \label{eq:chmpyr}
C_{\text{HM},\text{Pyr}}(t) &  =  & \frac{\mathbb{E}[|D_{\text{ag}}|]}{k_t} .
\eea
The problem thus reduces to finding expectation of $|D_{\text{ag}}|$. For every $\underline{u} \in \mathbb{F}_2^t, \underline{v} \in \mathbb{F}_2^{t-w_H(\underline{u})}, \ell \in [\mu(\underline{u},\underline{v})]$, define the random variable
\bean
M_{\ell}(\underline{u}, \underline{v}) & = & |\{ j \in [n_e] \mid \underline{\epsilon}_j  \in B_{\ell}(\underline{u}, \underline{v}) \}|
\eean as the number of rows subjected to an erasure pattern in $B_{\ell}(\underline{u}, \underline{v})$. Next we define
\bean 
M(\underline{u}, \underline{v}) & = & \max_{\ell}  M_{\ell}(\underline{u}, \underline{v}) .
\eean
The strategy ${\cal A}_t$ attempts to aggregate code symbols over $M(\underline{u}, \underline{v})$ rows for every observed type $(\underline{u}, \underline{v})$. Therefore, if $M(\underline{u},\underline{v}) = 0$, then it does not impact the value of $|D_{\text{ag}}|$, but if $M(\underline{u}, \underline{v}) > 0$, aggregation will happen over those rows and columns given by  $A(\underline{u})$. The size of aggregated set of code symbols for each $(\underline{u}, \underline{v})$ depends on the dimension accumulated by both the unaffected local codes and the overwhelmed local codes. Let $\mathbbm{1}(x)$ be an indicator function, i.e., $\mathbbm{1}(x) = 1$ if $x \neq 0$, and $\mathbbm{1}(x) = 1$ if $x = 0$. We can express $D_{\text{ag}}$ and compute its expectation as given below:
\bea \label{eq:dagsize} 
|D_{\text{ag}}| & = & \sum_{\underline{u} \in \mathbb{F}_2^t} \sum_{\underline{v} \in \mathbb{F}_2^{t-w}}  \mathbbm{1}(M(\underline{u}, \underline{v})) \left[ \sum_{i\in [t]\setminus \textsl{Supp}(\underline{u})} k_t(i) M(\underline{u}, \underline{v}) + \sum_{i \in \textsl{Supp}(\underline{u})} k_t(i) \right] + \left( n_e- \sum_{\underline{u} \in \mathbb{F}_2^t} \sum_{\underline{v} \in \mathbb{F}_2^{t-w}} M(\underline{u}, \underline{v}) \right) k_t \\
\nonumber \mathbb{E}[|D_{\text{ag}}|] & = & \sum_{\underline{u} \in \mathbb{F}_2^t} \sum_{\underline{v} \in \mathbb{F}_2^{t-w}} \left[ \sum_{i\in [t]\setminus \textsl{Supp}(\underline{u})} k_t(i) \mathbb{E}[M(\underline{u}, \underline{v})] + \sum_{i \in \textsl{Supp}(\underline{u})} k_t(i)\Pr(M(\underline{u}, \underline{v})\geq 1)  \right] + \\  
\label{eq:exdagsize} & & \hspace{8cm} \left( n_e- \sum_{\underline{u} \in \mathbb{F}_2^t} \sum_{\underline{v} \in \mathbb{F}_2^{t-w}} \mathbb{E}[M(\underline{u}, \underline{v})] \right) k_t 
\eea  
In order to compute the expression, it is required to determine both $\mathbb{E}[M(\underline{u}, \underline{v})]$ and $\Pr(M(\underline{u}, \underline{v})\geq 1)$ for every $(\underline{u}, \underline{v})$. Following the approach in \cite{PrRPA2020}, we identify the random variable $M(\underline{u}, \underline{v})$ with a quantity associated with well-known balls-and-bins experiment. Every row of $(n_e \times n_h)$-codeword is identified as an unlabeled ball, and each erasure pattern in ${\cal E}$ as a labelled bin. The random experiment is to throw these $n_e$ unlabeled balls at ${n_h \choose s}$ labeled bins. The number of balls in a bin, referred to as bin-size, is the number of rows subjected to the same erasure pattern. Without loss of generality, let us restrict focus to a single type, i.e., we focus on to bins associated with ${S_{(\underline{u},\underline{v})}}$. Since we are interested in number of rows subjected to equivalent erasure patterns, we will bunch up multiple (real) bins to form an artificial bin. As the bunching factor remains the same for a given type, we bunch up a fixed number of real bins. It follows that $M(\underline{u}, \underline{v})$ is maximum over bin-sizes of all artificial bins of a fixed type. Thus the problem reduces to the following: Throw $r$ unlabeled balls in $n$ labeled bins uniformly at random. The first $bm$ bins are bunched up to form $m$ artificial bins each constituting of $b$ real bins. Let $X_i,  i = 1,2,\ldots, n-bm+m$ denote the bin-sizes of all bins, and let $Z=\max_{i=1,2,\ldots,m}X_i$. Let us define $\rho(n,r,m,b) = \mathbb{E}[Z]$ and $\phi(n,r,m,b)=\Pr(Z \geq 1)$. Then we have
\bea \label{eq:exm}
\mathbb{E}[M(\underline{u},\underline{v})] & = & \rho\left({n_h \choose s},n_e,\frac{|S_{(\underline{u},\underline{v})}|}{\beta(\underline{u},\underline{v})},\beta(\underline{u},\underline{v})\right) \\
\label{eq:prm} \Pr[M(\underline{u},\underline{v}) \geq 1] & = & \phi\left({n_h \choose s},n_e,\frac{|S_{(\underline{u},\underline{v})}|}{\beta(\underline{u},\underline{v})},\beta(\underline{u},\underline{v})\right)
\eea
Both $\rho(n,r,m,b)$ and $\phi(n,r,m,b)$ are determined in App.~\ref{app:bsb}. Substituting \eqref{eq:exm}, \eqref{eq:prm} back in \eqref{eq:exdagsize} and then to \eqref{eq:chmpyr}, we obtain an expression for communication cost at helper-nodes for pyramid scheme $({\cal C}_t,{\cal A}_t)$. Thus we have proved the theorem:
\bthm Let $(n_e,n_h,s)$ be given. The pyramid scheme $({\cal C}_t,{\cal A}_t)$ for coded aggregation of gradient achieves the communication cost-pair  $(C_{\text{EH},\text{Pyr}}(t),C_{\text{HM},\text{Pyr}}(t))$ as given in \eqref{eq:cehpyr}, \eqref{eq:chmpyr} for $2 \leq t \leq \lfloor \frac{n_h-s+1}{3} \rfloor$.
\ethm

\section{An Achievable Tradeoff Between $C_{\text{EH}}$ and $C_{\text{HM}}$ \label{sec:tradeoff}}

The communication cost-pairs achieved by the pyramid scheme combined with that of AMC scheme permit to achieve a tradeoff between $C_{\text{EH}}$ and $C_{\text{HM}}$. 

\subsection{$C_{\text{HM}}$ of AMC \label{sec:chmamc}}

In \cite{PrRPA2020}, authors derived an upperbound on the communication cost at helpers $C_{\text{HM}}$ of the AMC scheme. With no further modification, their bound can be converted to an equality with the help of $\rho(\cdot)$ function. Thus we obtain the following theorem.
\bthm Let $(n_e,n_h,s)$ be given. The communication cost-pair $(C_{\text{EH},\text{AMC}}, C_{\text{HM},\text{AMC}})$ of AMC scheme is given by
\bean
C_{\text{EH},\text{AMC}} & = & \frac{n_h}{n_h-s} \\
C_{\text{HM},\text{AMC}} & = & n_e+1- \rho\left({n_h \choose s},n_e,{n_h \choose s},1\right)
\eean 
\ethm

\subsection{Illustration of Tradeoff with An Example \label{sec:egtradeoff}}

For practical reasons, it is desirable to scale $n_h$ logarithmically with $n_e$ and to keep $s=\alpha n_h$ as a fraction of $n_h$. Following this rationale, we choose example parameters $(n_e,n_h,s)=(2048,16,5)$ to illustrate the tradeoff in Fig.~\ref{fig:tradeoff}.

\section{Improved AMC considering $m$ maximas}

In the AMC scheme \cite{PrRPA2020}, the master finds the maximum number of matching rows in the erasure matrix. Within these matching rows, the helper nodes can aggregate and transmit the aggregated vector to the master node. For the remaining rows, the helper nodes transmit the received messages without any aggregation to the master. We observe that the aggregation can be performed at the helper nodes not just for the maximum number of matching rows, but even for second maximum, third maximum and so on. To illustrate this, consider the system model with $n_e =5, n_h =4$ and $s=1$. Let the observed erasure matrix be 
\bean \mathcal{M}_{E} &= & \begin{bmatrix}
0 & 0 & 0 & 1 \\
0 & 0 & 0 & 1 \\
0& 0 & 0 & 1 \\
1 & 0 & 0 & 0 \\
1 & 0 & 0 & 0 \\
\end{bmatrix} .
\eean 
Since the maximum matching rows is three, as per the AMC scheme, the helper nodes aggregates the first three rows and transmit the aggregated gradient. For the last two rows, it transmits all the received messages directly. However, the helper nodes can aggregate the messages in the last two rows and reduce the cost of transmission at helper nodes. As per the AMC scheme, for the observed erasure matrix $\mathcal{M}_{E}$ the communication cost at the helper nodes is $3$. By considering the second maxima also, the communication cost at the helper nodes can be reduced to $2$. 

The idea is to generalize the AMC scheme by considering more than one set of maximum matching rows. We propose a scheme in which the helper nodes consider $m$ maximas for some integer $m \geq 1$. For $m=1$ the proposed scheme reduces to the AMC scheme. Let $M_i, i \in [m]$ denote the number of $i$-th maximum set of matching rows. The helper nodes perform aggregation at each of these matching sets and transmits aggregated gradient vectors. For $i \in [m]$, the helper nodes can aggregate $M_i$ rows and transmit the aggregated partial gradients. The communication cost incurred for the transmission of aggregated partial gradient for each set of matching rows is $(n_h - s) \frac{1}{n_h-s} = 1$. For the remaining rows, similar to the AMC scheme, the helper nodes transmit the received messages without any aggregation to the master. Hence, the communication cost incurred at the helper nodes is equal to $\left(n_e - \sum_{i=1}^{m} M_i\right)(n_h-s)\frac{1}{n_h-s} = \left(n_e - \sum_{i=1}^{m} M_i\right)$. Hence, for a particular erasure matrix $\mathcal{M}_{E}$ the communication cost at the helper nodes for the scheme considering $m$ maximas is 
$$
n_e - \sum_{i=1}^{m} M_i + m.
$$ 
The normalized communication load is obtained by averaging over all erasure patterns. Hence, we have, $$C_{HM} = n_e - \sum_{i=1}^{m} \mathbb{E} [M_i] + 1. $$

\begin{figure}[h]
	\begin{center}
		\includegraphics[scale=0.4]{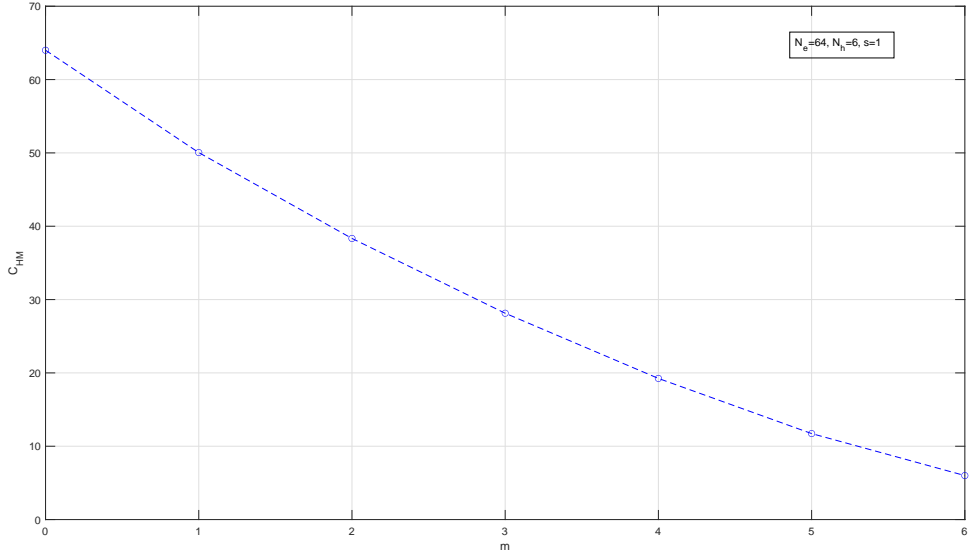}
		\caption{$C_{HM}$ values corresponding to $m \in [6]$ for $(n_e,n_h,s) = (64,6,1)$. }
		\label{fig:AMC_ne64_nh4_m}
	\end{center}
\end{figure}

\begin{figure}[h]
	\begin{center}
		\includegraphics[scale=0.4]{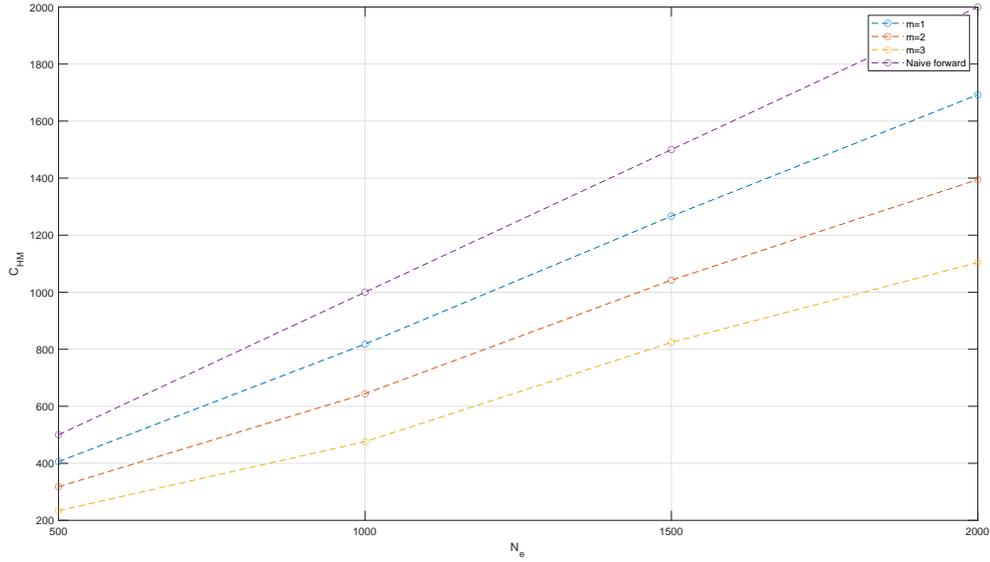}
		\caption{Comparison of variation of $C_{HM}$ with $n_e$ for different values of $m$.}
		\label{fig:AMC_CHM_vs_N_e}
	\end{center}
\end{figure}

For a system with $n_e = 64$, $n_h = 6$ and $s=1$, using Monte-Carlo technique, the $C_{HM}$ values for various values of $m$ are obtained and plotted in Fig. \ref{fig:AMC_ne64_nh4_m}. It can be observed from Fig. \ref{fig:AMC_ne64_nh4_m} that as $m$ increases, more and more aggregation opportunities arise and hence a reduction in $C_{HM}$ is obtained. In the graph $m=0$ indicates the naive scheme in which the helper nodes transmits all the received messages directly to the master node without any aggregation. We performed simulation to compute the normalized communication cost for different values of $n_e$, keeping $n_h = \lfloor \text{log}(n_e) \rfloor$, and $s = 0.2 * n_h$. The communication cost at the helper nodes reduces with increasing values of $m$ as evident from Fig. \ref{fig:AMC_CHM_vs_N_e}. 

\section{A greedy approach to ARC}

We begin with illustrating the ARC scheme with an example. 
\begin{example}
	Consider a system with $n_e = 4$ client nodes and $n_h = 6$ helper nodes with at most $s=2$ straggling links between each client and helper nodes. The client node partitions the gradient vector into $k = \frac{n_h}{s+1} = 2$ parts and uses a generator matrix 
	\bean 
	G & = & 
	\begin{bmatrix}
	1 & 0 & 1 & 0 & 1 & 0\\
	0 & 1 & 0 & 1 & 0 & 1
	\end{bmatrix}.
\eean Let the observed erasure matrix be 
\bean \mathcal{M}_E & = & \begin{bmatrix}
	0 & 1 & 1 & 0 & 0 & 0 \\
	1 & 0 & 0 & 0 & 0 & 1 \\
	1 & 0 & 1 & 0 & 0 & 0 \\
	0 & 1 & 0 & 1 & 0 & 0 
	\end{bmatrix} .
\eean 
The observed erasure pattern is also depicted in Fig. \ref{fig:erasureARC}. Following the aggregation strategy employed in ARC, it can be observed that all the helper nodes needs to transmit the messages and hence the communication cost at the helper nodes for this particular scenario is $s+1 = 3$. 
	
	\begin{figure}
		\centering 
		\scalebox{.9}{
			\begin{tabular}{|c|c|c|c|c|c|c|c|c|} 
				\hline
				& $H_1$ & $H_2$ & $H_3$ & $H_4$ & $H_5$ & $H_6$ \\
				\hline
				$E_1$ & \boxit{0.2in}{0.07in}$g_{11}$ &  \xcancel{$g_{12}$} &  \xcancel{$g_{11}$} &  \boxit{0.2in}{0.07in}$g_{12}$ &  $g_{11}$ & $g_{12}$ \\
				\hline
				$E_2$ & \xcancel{$g_{21}$} &  \boxit{0.2in}{0.07in}$g_{22}$ &  \boxit{0.2in}{0.07in}$g_{21}$ &  $g_{22}$ &  $g_{21}$ & \xcancel{$g_{22}$}  \\
				\hline
				$E_3$ & \xcancel{$g_{31}$} &  \boxit{0.2in}{0.07in}$g_{32}$ &  \xcancel{$g_{31}$} &  $g_{32}$ &  \boxit{0.2in}{0.07in}$g_{31}$ & $g_{32}$ \\
				\hline
				$E_4$ & \boxit{0.2in}{0.07in}$g_{41}$ &  \xcancel{$g_{42}$} &  $g_{41}$ &  \xcancel{$g_{42}$} &  $g_{41}$ & \boxit{0.2in}{0.07in}$g_{42}$ \\
				\hline
		\end{tabular} }
		\caption{\footnotesize Aggregation strategy of the ARC scheme for $(n_e=4,n_h=6,s=2)$ .}
		\label{fig:erasureARC}
	\end{figure} 
	\label{eg:ExampleARC}
\end{example}

The principal idea of ARC is to repeat the components of the computed gradient to helper nodes without coding. Therefore the gradient aggregation at the helper nodes for ARC can be viewed as a set cover problem \cite{Cormen}. Having recognized that, we can employ a greedy algorithm to solve it. We show by simulations the greedy algorithm performs better than the existing aggregation strategy followed in \cite{PrRPA2020}. 

\subsection{Greedy approach}
\label{subsec:GA}
The objective of the master is to compute the total gradient  $
\underline{g}_D =  \left[\sum_{i=1}^{n_e} \underline{g}_{i,1} \ \sum_{i=1}^{n_e} \underline{g}_{i,2} \ \cdots \ \sum_{i=1}^{n_e} \underline{g}_{i,k}\right] .
$ Let $\mathcal{U} = \bigcup_{\substack{i \in [n_e] \\ j \in [k]}} \underline{g}_{i,j}$ denote a set of gradient vectors. Each helper node $H_i$ receives a subset of partial gradients denoted by $\mathcal{S}_i$. The master nodes wants each element in $\mathcal{U}$ to be aggregated and transmitted by one helper node. The master can then combine the transmissions suitably to find the total gradient $\underline{g}_D$. The master node effectively wants to find a cover of the universal set of gradients $\mathcal{U}$ using minimum number of sets from $\mathcal{S}_i, i \in [n_h]$. The minimum number is desired since the communication cost at the helper nodes depends on the number of helper nodes participating in the aggregation phase. Hence we can model the aggregation phase of ARC to a set cover problem with universal set $\mathcal{U}$ and subsets $\mathcal{S}_i, i \in [n_h]$. The set covering problem is an NP-complete problem with a heuristic greedy algorithm \cite{Cormen}. The greedy algorithm for the aggregation phase in ARC scheme is described in Algorithm \ref{alg:greedyARC}.

\begin{algorithm}[H]
	\caption{Greedy algorithm}
	\begin{algorithmic}[1]
	\State Input: $\mathcal{U}$ and $\mathcal{S}_i, i \in [n_h]$
	\State Initialize $U \leftarrow \mathcal{U}, C \leftarrow \phi$
	\While {$U \neq \phi$} 
	 	\State Select an $i \in [n_h]$ such that $\vert \mathcal{S}_i \cap U \vert$ is maximum
	 	\State $\mathcal{A}_i = \mathcal{S}_i \cap U$ 
	 	\State $C \leftarrow C \cup i$
	 	\State $U \leftarrow U \setminus \mathcal{S}_i$ 
	 	
	 	\EndWhile
	 	\State For every $i \in C$, the master node asks helper node $H_i$ to aggregate the partial gradients in $\mathcal{A}_i$ and transmit.
        	\end{algorithmic}
        	\label{alg:greedyARC}
    \end{algorithm}

\begin{example}
For the system considered in Example \ref{eg:ExampleARC} and the erasure patter as described in Fig. \ref{fig:erasureARC}, the greedy algorithm outputs $C=\{5,6,2\}, \mathcal{A}_5 = \{\underline{g}_{1,1}, \underline{g}_{2,1},\underline{g}_{3,1}, \underline{g}_{4,1} \}$, $\mathcal{A}_6 = \{ \underline{g}_{1,2}, \underline{g}_{3,2}, \underline{g}_{4,2} \}$ and $\mathcal{A}_{2} = \{ \underline{g}_{2,2} \}$. The communication cost at the helper nodes for this particular scenario is equal to $1.5$. 
\end{example}

\subsection{Simulation Results}
\begin{figure}[h]
	\begin{center}
		\includegraphics[scale=0.4]{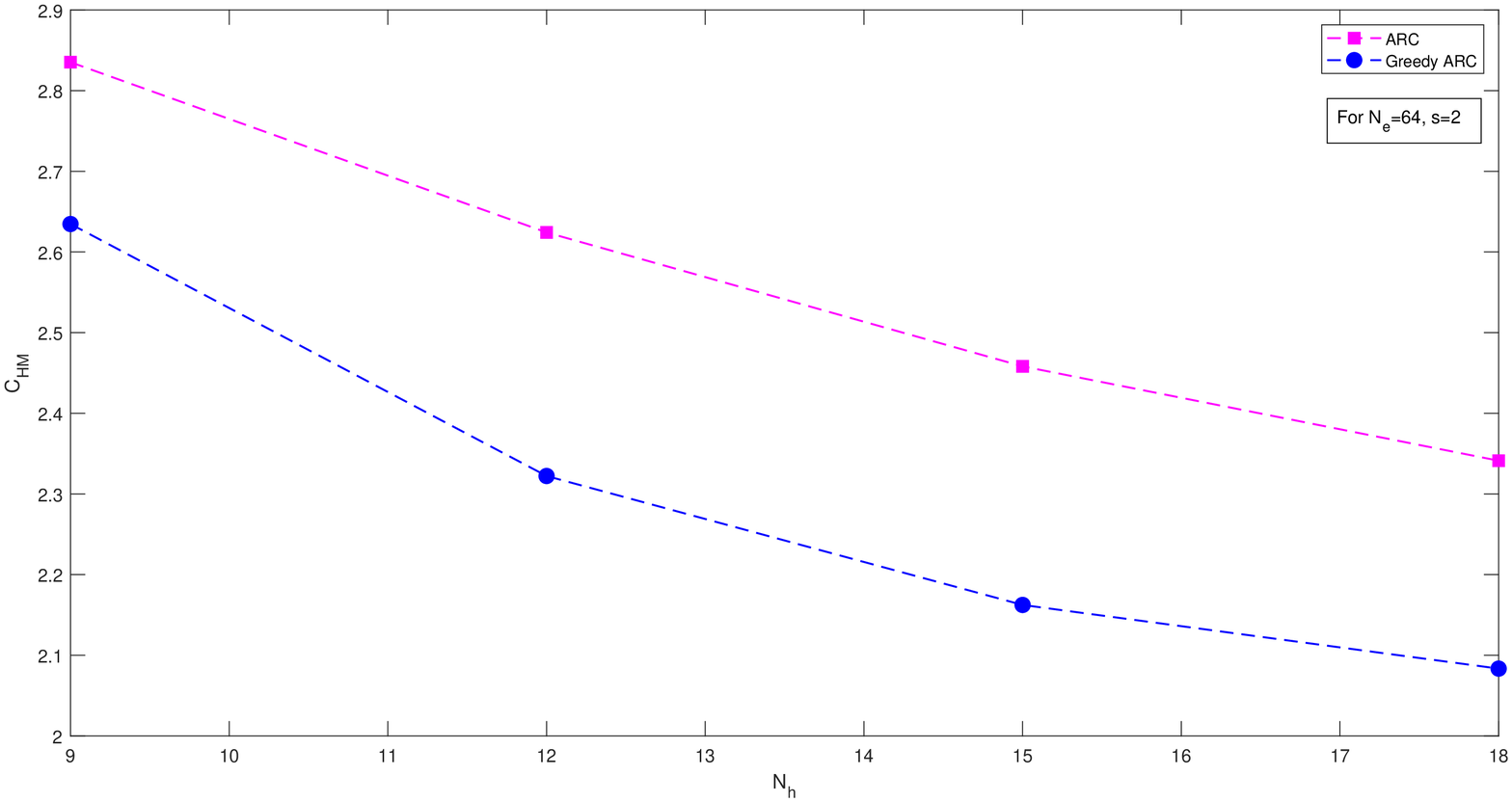}
		\caption{Comparison of $C_{HM}$ values of ARC scheme and the proposed greedy approach for $n_e = 64, s=2$ and for different values of $n_h$ }
		\label{fig:GreedyARCSim}
	\end{center}
\end{figure}

The example described in Section \ref{subsec:GA} considers only one particular erasure pattern. Recall that the normalized communication load $C_{HM}  =  \ \frac{1}{|\Omega(s)|} \sum_{\epsilon \in \Omega(s)} \sum_{j=1}^{n_h} \frac{m_{j,\epsilon}}{k}.$ We consider a system with $n_e = 64$ client nodes and $s=2$ straggling links per client node. The $C_{HM}$ value for both greedy approach and the existing strategy is computed using Monte-Carlo technique by averaging over a large number of iterations.  The performance improvement of the greedy approach over the existing ARC scheme can be observed from Fig. \ref{fig:GreedyARCSim}. The $C_{HM}$ values are plotted for different $n_h$ values. 

\bibliographystyle{IEEEtran}
\bibliography{cag}

\appendices

\section{Maximum Bin-Size With Bunching \label{app:bsb}}

Suppose $r$ unlabeled balls are thrown at $n$ labeled bins uniformly at random. The first $bm$ bins, $bm \leq n$ are {\em bunched} to form $m$ artificial bins each consisting of $b$ real bins. Now there are a total of $\nu=n-bm+m$ bins. Let random vector $\underline{X} = [X_1 \ X_2 \ \cdots \ X_{\nu}]$ denote the number of balls in each bin. When $k_1 + k_2 + \cdots + k_{\nu} = r$, 
\bean
\Pr (\underline{X} = (k_1,k_2,\ldots, k_{\nu}))  =  \frac{r!}{k_1! k_2! \cdots k_{\nu}!} \left(\frac{b}{n}\right)^{k_1+k_2+\cdots + k_m} \left(\frac{1}{n}\right)^{k_{m+1}+k_{m+2}+\cdots + k_{\nu}}
\eean 
We define $M = \max_{i = 1,2,\ldots, m} X_i$. We are interested in determining both $\rho(n,r,m,b)=\mathbb{E}[M]$ and $\phi(n,r,m,b)=\Pr[M \geq 1]$. Since $M$ is a non-negative integer valued random variable,
\bea
\mathbb{E}[M] & = & \sum_{i=1}^{r} \Pr(M \geq i) \nonumber \\
& = & \sum_{i=1}^{r} \Pr( \cup_{j=1}^m \{X_j \geq i \}) \label{eq:ex1}
\eea
Define $E_{ij} \ = \ \{X_j \geq i \}$ and apply inclusion-exclusion principle to obtain $\rho(n,r,m,b) \ = \ \sum_{i=1}^{r} \sum_{\ell=1}^m A_{i\ell}$ where $A_{i\ell}$ is given in \eqref{eq:aiell}. 
\bea
\label{eq:aiell}A_{i\ell} & = & (-1)^{\ell-1} \cdot {m \choose \ell} \cdot \sum_{\substack{k_1\geq i, k_2\geq i, \ldots, k_{\ell}\geq i \\ \sum_{i\in [\ell]}k_i \leq r}} \frac{r!}{k_1! k_2!\cdots k_{\ell}!(r-\sum_{i\in [\ell]}k_i)!} \left(\frac{b}{n}\right)^{\sum_{i\in [\ell]}k_i} \left(1- \frac{\ell b}{n}\right)^{r-(\sum_{i\in [\ell]}k_i)}
\eea
Then it follows that
\bea
\rho(n,r,m,b) & = & \sum_{\ell=1}^{m}\sum_{i=1}^{\left\lfloor\frac{r}{\ell}\right\rfloor} A_{i\ell} \label{eq:rho} \\
\phi(n,r,m,b) & = & \sum_{\ell=1}^{m} A_{1\ell} \label{eq:phi}
\eea

\end{document}